\documentclass[conference,10pt]{IEEEtran}
\usepackage{subfig}
\usepackage{float}
\usepackage{amsfonts, bm}
\usepackage{amssymb}
\usepackage{color,graphicx}
\usepackage{cite}
\usepackage[cmex10]{amsmath}
\usepackage{amsopn}
\usepackage{multirow}
\usepackage{bigstrut}
\usepackage{microtype}
\usepackage{geometry}
\usepackage{algorithmic}
\usepackage[ruled]{algorithm2e}

\IEEEoverridecommandlockouts

\def\gap{1.0ex}
\abovedisplayskip\gap
\belowdisplayskip\gap
\abovedisplayshortskip\gap
\belowdisplayshortskip\gap

\allowdisplaybreaks

\newtheorem{proposition}{Proposition}

\newtheorem{remark}{Remark}

\begin{document}
	
	\sloppy
	
	\title{Design and Optimization of Cooperative Sensing With Limited Backhaul	Capacity}
	\author{Wenrui Li, Min Li, An Liu and Tony Xiao Han ~\thanks{
			\par Wenrui Li, Min Li and An Liu are with College of Information Science and Electronic Engineering, Zhejiang University, and also with Zhejiang Provincial Key Laboratory of Information Processing, Communication and Networking, Hangzhou 310027, China (Emails: \{12031075, min.li, anliu\}@zju.edu.cn). Tony Xiao Han is with Huawei Techologies Co., Ltd., Shenzhen 518129, China (Email: tony.hanxiao@huawei.com). This work was supported in part by National Natural Science
			Foundation of China under Grant 62271440, the Fundamental Research
			Funds for the Central Universities 226-2022-00195, and a Research Project
			funded by Huawei Technologies Co., Ltd.}}
	\normalsize
	\maketitle
	
	\begin{abstract}
		This paper introduces a cooperative sensing framework designed for integrated sensing and communication cellular networks. The framework comprises one base station (BS) functioning as the sensing transmitter, while several nearby BSs act as sensing receivers. The primary objective is to facilitate cooperative target localization by enabling each receiver to share specific information with a fusion center (FC) over a limited capacity backhaul link. To achieve this goal, we propose an advanced cooperative sensing design that enhances the communication process between the receivers and the FC. Each receiver independently estimates the time delay and the reflecting coefficient associated with the reflected path from the target. Subsequently, each receiver transmits the estimated values and the received signal samples centered around the estimated time delay to the FC. To efficiently quantize the signal samples, a Karhunen-Lo\`eve Transform coding scheme is employed. Furthermore, an optimization problem is formulated to allocate backhaul resources for quantizing different samples, improving target localization. Numerical results validate the effectiveness of our proposed advanced design and demonstrate its superiority over a baseline design, where only the locally estimated values are transmitted from each receiver to the FC.
	\end{abstract}


	\section{Introduction}\label{Introduction}
	\par Recently, Integrated Sensing and Communication (ISAC) has garnered significant attention from both industry and academia~\cite{ISAC4,ISAC0}. This emerging concept involves the sharing of hardware and/or resources between sensing and communication systems, allowing them to mutually benefit from the integrated design. On one hand, sensing contributes to more accurate channel estimation in communication by providing environmental awareness. On the other hand, communication enables the exchange of sensing information, thereby enhancing the performance of sensing. As a result, ISAC is expected to find applications in various fields within future 6G systems, including autonomous vehicles, unmanned aerial vehicles (UAVs), extended reality (XR), and more~\cite{ISAC4,ISAC0,liu2022information}.
	
	\par There are two fundamental designs for ISAC: the resource-sharing design and the joint design. In the resource-sharing design, system resources (such as time/frequency resources) are partitioned, with dedicated resources used for sensing and communication separately. Conversely, in the joint design, sensing and communication functions are integrated using the same signal waveform and shared resources. While the resource-sharing design is generally suboptimal for ISAC, it remains widely used in practical ISAC systems due to its simplicity and better compatibility with existing communication systems.
	
	\par This paper focuses on the resource-sharing ISAC design and investigates a cooperative sensing framework within the context of ISAC cellular networks. In this framework, one base station (BS) acts as the sensing transmitter, while several nearby BSs serve as sensing receivers, as depicted in Fig.~\ref{Fig:1}. The sensing transmitter sends the sensing signal on dedicated resources, and each sensing receiver captures the signal reflected by the target and communicates with a fusion center (FC) through a backhaul link with limited capacity, similar to cloud radar systems\cite{cooperative-radar1,cooperative-radar2,cooperative-radar4,cooperative-radar5}. The FC's objective is to estimate the location of the target. The main challenge lies in designing an appropriate quantization scheme at each receiver to optimize the cooperative sensing performance while adhering to the constraints imposed by the limited backhaul capacity.

	\par Several relevant studies in this area have been documented in the literature. In~\cite{cooperative-radar1,cooperative-radar2}, the standard approach of modeling the effect of quantization by means of an additive quantization noise was adopted, and the problem of determining the optimum transmit code vector and quantization error covariance matrices was studied to optimize the target detection performance at the FC. However, their treatment of quantization relied on rate-distortion theory and lacked a concrete practical scheme. Another work, \cite{cooperative-radar4}, adopted a practical uniform quantizer to quantize measurements from different receivers. They derived the corresponding quantization error and its Gaussian approximation to evaluate the impact of finite backhaul capacity on target localization performance. Nonetheless, directly sending uniformly-quantized received measurements to the FC is generally suboptimal unless the backhaul capacity is sufficiently large.

	\par To address these challenges and efficiently utilize backhaul resources, we propose an advanced design in this paper for receiver-FC communication. Initially, the $n$-th receiver transmits its estimated values of the time delay $\tau _n$ and reflecting coefficient $\alpha _n$, related to the reflected path from the target to the $n$-th receiver. Then, each receiver transmits limited amount of the received sensing signal sampled around the estimated time delay to the FC accounting for the limited backhaul capacity constraint. To achieve this efficiently, we propose a quantization scheme based on a combination of Karhunen-Lo\`eve Transform (KLT)~\cite{KLT2001} and Lloyd scalar quantization, enabling the transmission of very few sampled signals around the estimated time delay from each receiver to the FC. To optimize backhaul capacity allocation for the Lloyd quantization segment, we present an iterative algorithm to solve the problem effectively. Additionally, when no quantization bits are allocated to the signal samples, the FC estimates the target's parameters solely based on the estimated values, which we refer to as the baseline design. Numerical results validate the effectiveness of our advanced design and demonstrate its significant improvement in target estimation accuracy compared to the baseline design.
	
	
	\par \textit{Notations}: Notations $\textrm{Re}(x)$ and $\textrm{Im}(x)$ denote the real and imaginary part of a complex number $x$, respectively. Notations ${\bf A}^T$ and ${\bf A}^{-1}$ denote the transpose and inverse of matrix ${\bf A}$, respectively. ${\bf A}_{i,j}$ denotes the element at the $i$th row and the $j$th column of ${\bf A}$. Notation $|{\bf x}|$ denotes the $l_1$-norm of vector ${\bf x}$, while $\left|{\bf x}\right|_2$ denotes the $l_2$-norm of vector ${\bf x}$. Notation $\mathcal{R}$ is used to denote the set of real numbers, respectively. ${\bf{I}}_{j}$ represents an $j \times j$ dimensional identity matrix. Notation $\textrm{diag}({\bf A})$ collects the diagonal elements of matrix ${\bf A}$, while $\textrm{diag}({\bf a})$ is to create a diagonal matrix with vector ${\bf a}$ being the diagonal elements. Finally, ${\mathcal{CN}}(a,b)$ denotes the complex Gaussian distribution with mean $a$ and variance $b$, respectively.
	
	\section{System Model and Problem Statement} \label{sec:model}
	\begin{figure}[t]
		\centering
		\includegraphics[width=0.35\textwidth]{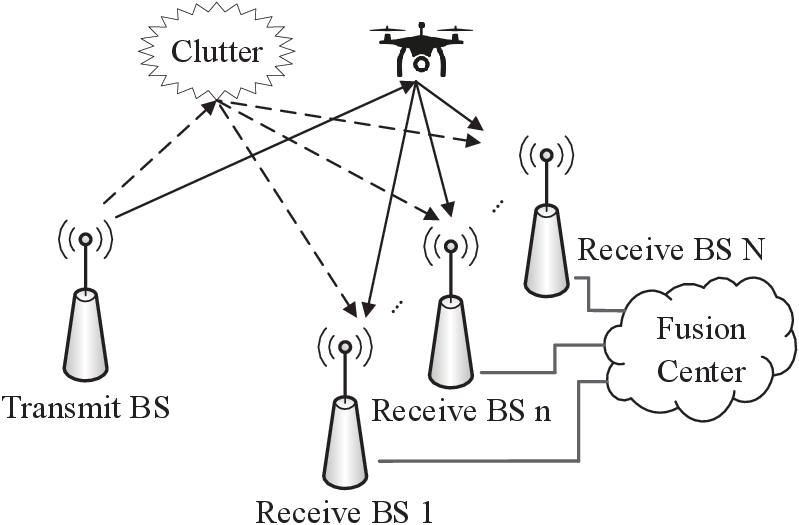}\caption{An illustration of the cooperative sensing network considered.}
		\label{Fig:1}
		\vspace*{-1.5\baselineskip}
	\end{figure}
	\par We consider a cooperative sensing scenario in the context of cellular networks, where one BS acts as sensing transmitter while the other $N$ BSs act as sensing receivers, and each receiver is connected to a FC with a finite capacity backhaul link, as shown in Fig.~\ref{Fig:1}. The sensing signal is sent by the transmitter, reflected by the target, and then received by the $N$ receivers. After processing the received signals in certain way and producing some local results, each receiver quantizes the local results and then forwards to the FC subject to limited backhaul capacity. Finally, the FC estimates the target's parameters based on the quantized local results from all receivers. The backhaul capacity for sending the local results from the $n$-th receiver to the FC is assumed to be $C_n$ bits within the processing interval.
	
	\par The transmitter is located at position $({x^t},{y^t})$, and the $n$-th receiver is located at position $(x_n^r,y_n^r)$, which are fixed and assumed to be known a priori. The lowpass equivalent of the signal transmitted from the transmitter is $\sqrt E s(t)$, where $E$ is the transmitted energy, and the waveform $s(t)$ satisfies $\int_{T_c} {{{\left| {s(t)} \right|}^2}dt}  = 1$, with $T_c$ being the signal duration time.
	
	\par Assume that the target has already been detected, but its location $({x},{y})$ is deterministic unknown and to be estimated. The sensing signal transmitted might traverse a primary line-of-sight (LOS) path along with non-line-of-sight (NLOS) paths before reaching the $n$th receiver. Normally, only LOS path is exploited for localization purpose~\cite{gezici2005localization,he2010noncoherent}, while the NLOS paths can be treated as part of clutter. Similar to~\cite{he2010noncoherent}, one can assume that the noise-plus-clutter corresponding to each dominant path is a temporally white, zero-mean complex Gaussian random process. Then the signal received at the $n$-th receiver is 
	\begin{equation} \label{r1}
		{r_n}(t) = \sqrt E {\rho _n} {\xi _n} s(t - {\tau _n})  + {w_n}(t),
	\end{equation}
	where ${\xi _n}$ is the reflection coefficient for the $n$th receiver, which is assumed to be a random complex variable, with unknown deterministic amplitude and random phase uniformly distributed between $0$ and $2\pi$, and ${w_n}(t)\sim {\mathcal{CN}}(0,\sigma_n^2)$ is the complex Gaussian effective noise term at the $n$-th receiver. In addition, ${\tau _n}$ and ${\rho _n}$ are the time delay and the path-loss coefficient corresponding to the $n$-th receiver respectively, which are both assumed to be unknown deterministic variables. In particular, the time delay ${\tau _n}$ is a function of the target's location $({x},{y})$ as:
	\begin{equation} \label{delay}
		{\tau _n} = \frac{{\sqrt {{{({x^t} - x)}^2} + {{({y^t} - y)}^2}}  + \sqrt {{{(x_n^r - x)}^2} + {{(y_n^r - y)}^2}} }}{c},
	\end{equation}
	where $c$ denotes the speed of light.
	
	\par
	Defining ${\alpha _n}={\rho _n} {\xi _n}$ as an effective reflecting coefficient, the discretized received signal at the $n$-th receiver becomes
	\begin{align} \label{r2}
		{r_n}(k{T_s}) &= \sqrt E {\alpha _n} s(k{T_s} - {\tau _n}) + {w_n}(k{T_s}),  k = 1,2, \cdots ,K,
	\end{align}
	where ${T_s}$ is the sampling period, $K$ is the number of samples and ${w_n}(k{T_s}) \sim {\mathcal{CN}}{(0,{\sigma _n^2})}$.
	
	\par
	Define a parameter vector ${\bm{\theta }}$ that collects the parameters of interest to be estimated
	\begin{equation} \label{eqtheta}
		{\bm{\theta }}  = {[x,y]^T},
	\end{equation}
	which is the coordinate of the location of the target. The focus of this article is to design proper local processing and quantization strategies at each receiver  so that the performance of estimating the target parameter vector ${\bm{\theta}}$ at the FC can be optimized subject to the backhaul resource given.
	
	\section{An Advanced Cooperative Sensing Design} \label{sec:advanced:design}
	\par In this section, we propose an advanced cooperative sensing design that consists of three components: \textit{i}) the Maximum Likelihood (ML) estimator of individual time delay and reflecting coefficient at each receiver; \textit{ii}) the quantization of the received signal samples at each receiver; and \textit{iii}) the ML estimator of the target's location at the FC.  Each component is discussed in details in the following subsections.
	
	\subsection{ML Estimation of Individual Time Delay and Reflecting Coefficient}\label{sec:MLA}
	For the $n$-th receiver, the ML estimator of time delay ${\tau _n}$ and reflecting coefficient ${\alpha _n}$ can be expressed as:
	\begin{equation}
		[{{\hat \tau }_n},{\hat \alpha _n}] = \arg \mathop {\max }\limits_{{\tau _n},{\alpha _n}} \ln p({{\bf{r}}_n}|{\tau _n},{\alpha _n}), \label{MLofdelayalpha}
	\end{equation}
	where ${{\bf{r}}_n} = \{ {r_n}({k}{T_s})|{k} = 1, \cdots ,{K_0}\} $ and ${K_0}$ is large enough to ensure sufficiently wide interval to contain all the useful signal for estimating the time delay. From~\eqref{r2}, the posterior probability density function (pdf) of ${{\bf{r}}_n}$ against ${\tau _n}$, ${\alpha _n}$ in the log domain is given as
	\begin{align}
		&\ln p({{\bf{r}}_n}|{\tau _n},{\alpha _n}) = \nonumber\\
		&- \frac{1}{{\sigma _n^2}}\sum\limits_{{k} = 1}^{K_0} {{{\left| {{r_n}({k}{T_s}) - \sqrt E {\alpha _n} s({k}{T_s} - {\tau _n})} \right|}^2}}+D_0,  \label{pr_delayalpha}
	\end{align}
	where $D_0$ is a constant independent of ${\tau _n}$ and ${\alpha _n}$.
	
	\begin{remark}\label{assumption:gaussian:distributed}
		The estimated ${\hat \tau_n}$ is assumed to be statistically related to the true parameter $\tau_n$, following the model in~\cite{taumodel}:
		\begin{align} \label{Gaussiandelay}
			{\hat \tau _n} = {\tau _n} + {w_{\tau_n}},
		\end{align}
		where the variable ${w_{{\tau_ n}}}$ represents Gaussian noise with variance $\textrm{CRB}_{\tau_ n}$, the Cramer-Rao Bound (CRB) for estimating $\tau_n$. The CRB can be computed as follows:
		\begin{align} \label{CRBtaun0}
			\textrm{CRB}_{\tau _n} = \frac{1}{\frac{{2E}}{{\sigma _n^2}}{\left| {{\alpha _n}} \right|^2}\sum\limits_{{k} = 1}^{{K_0}} {{{\left[\frac{{\partial s(t)}}{{\partial t}}|_{t = kT_s  - \tau _n }\right]}^2}}}.
		\end{align}
		whose detailed proof is omitted due to space limitation.
	\end{remark}
	
	\subsection{Quantization of Received Signal Samples at Each Receiver}\label{sec:QuantizationB}
	\par We introduce an index set of samples at the $n$-th receiver as\\ ${\bf{K}}_{n}=\left\{k_{n}\in {\mathcal Z}|{\hat{\tau} _n} - \frac{{{T_d}}}{2} \le {k_{n}}{T_s} \le {\hat{\tau} _n} + \frac{{{T_d}}}{2}\right\}$, where $T_d$ is the length of the sampling interval, slightly larger than $T_c$ to tolerate estimation error. Then the corresponding received signal samples can be expressed as
	\begin{align}
		{r_{n}}({k_{n}}{T_s}) = \sqrt E {\alpha _n}&s({k_{n}}{T_s} - {\tau _n}) + {w_n}({k_{n}}{T_s}), {k_{n}} \in {{\bf{K}}_{n}}. \nonumber
	\end{align}
	The vector to be quantized is a $1\times 2K_{n}$ vector that collects the real and imaginary parts of the samples above, i.e.,
	\begin{align}
		{{\bf{r}}_n} = \left[\left \{ {\rm{Re}}({r_n}({k_n}{T_s})),{k_n} \in {{\bf{K}}_n} \right \}, \left \{ {\rm{Im}}({r_n}({k_n}{T_s})),{k_n} \in {{\bf{K}}_n} \right \}\right]. \nonumber
	\end{align}
	
	\par For efficient quantization of ${{\bf{r}}_n}$, it is crucial to exploit its statistics such as the joint distribution of vector components. However, it is difficult to characterize the exact distribution of ${\bf{r}}_{n}$ due to the complicate waveform distribution and the unknown random parameters involved. To make the analysis more tractable, we consider the first-order Taylor expansion of ${{\bf{r}}_n}$ at ${\hat \tau_n}$ as
	{\begin{align}
			{r'_{n}}({k_{n}}{T_s})=
			&\sqrt E {\alpha _n}\left[\frac{{\partial s(t)}}{{\partial t}}{|_{t = k_n{T_s} - {\hat \tau _n}}}({w_{{\tau _n}}}) + s(k_n{T_s} - {\hat \tau _n})\right] \nonumber\\
			&+ {w_n}({k_{n}}{T_s})\nonumber
	\end{align}}
	and form its associated $1\times 2K_{n}$ vector
	\begin{align}
		{{\bf{r}'}_{n}} = \left[\left \{ {\mathop{\rm Re}\nolimits} ({r'_{n}}({k_{n}}{T_s})),{k_{n}} \in {{\bf{K}}_{n}} \right\},
		\left\{ {\mathop{\rm Im}\nolimits} ({r'_{n}}({k_{n}}{T_s})),{k_{n}} \in {{\bf{K}}_{n}} \right\} \right] \nonumber
	\end{align}
	to approximate the statistics of ${\bf{r}}_{n}$. In this way, it is straightforward to show that ${{\bf{r}'}_{n}}$ is Gaussian distributed with mean and covariance matrix given in the following proposition.
	
	\begin{proposition}\label{proposition:first:order:approx}
		Signal vector ${{\bf{r}'}_{n}}$ is Gaussian distributed with its mean and covariance matrix given as:
		\begin{align}
			{{\overline{\bf{r}}}_{n}} = &\Big[\left \{\sqrt E \alpha _n^R s(k_n{T_s} - {\hat \tau _n}),{k_{n}} \in {{\bf{K}}_{n}}\right\}, \nonumber \\
			&\left \{ \sqrt E \alpha _n^I s(k_n{T_s} - {\hat \tau _n}),{k_{n}} \in {{\bf{K}}_{n}} \right\} \Big], \label{equ:mean} \\
			{{\bf{Q}}_{{{\bf{r}}_{n}}}}&= E \cdot \textrm{CRB}_{\tau_n} \cdot {\bf{q}}_{{{\bf{r}}_{n}}}^T{{\bf{q}}_{{{\bf{r}}_{n}}}}  + \frac{1}{2}\sigma _n^2{{\bf{I}}_{2{K_{n}}}}, \label{equ:Qr}
		\end{align}
		where ${{\bf{q}}_{{{\bf{r}}_{n}}}}=\Big[\left\{\alpha _n^R\frac{{\partial s(t)}}{{\partial t}}{|_{t = {k_{n}}{T_s} - {{\hat \tau }_n}}},{k_{n}} \in {{\bf{K}}_{n}}\right\},\nonumber \\ \left\{\alpha _n^I\frac{{\partial s(t)}}{{\partial t}}{|_{t = {k_{n}}{T_s} - {{\hat  \tau }_n}}},{k_{n}} \in {{\bf{K}}_{n}}\right\}\Big]$ is the $1 \times 2K_{n}$ derivative vector, and $ \alpha _n^R = {\mathop{\rm Re}\nolimits} ({{\alpha }_n})$, $\alpha _n^I = {\mathop{\rm Im}\nolimits} ({{\alpha }_n})$.
		
		\begin{IEEEproof} \label{proof:first:order:approx}
			The results follow from the definition of~${{\bf{r}'}_{n}}$ and the property of ${\hat \tau _n}$ as stated in Remark~\ref{assumption:gaussian:distributed}. The detailed proof is omitted due to space limitation.
		\end{IEEEproof}
	\end{proposition}
	
	\par It is also noted that when evaluating~\eqref{equ:mean} and \eqref{equ:Qr}, perfect knowledge of $\alpha_n$ is required. To relax this, we simply use the estimated values $\hat \alpha_n, n=1,\cdots,N$.
	
	\par With the above approximation, we now present a quantization scheme of ${\bf r}_n$. In particular, we propose to use Karhunen-Lo\`eve Transform (KLT) coding scheme that was previously shown to be mean-squared-error optimal for jointly Gaussian vector source compression~\cite{KLT2001}.
	
	\par The scheme involves two steps. Consider that the symmetric covariance matrix ${\bf{Q}}_{{\bf{r}}_{n}} = {\bf U} {\bf \Lambda} {\bf U}^T$ as a result of singular value decomposition, where diagonal matrix ${\bf \Lambda}$ collects all eigenvalues of ${\bf{Q}}_{{\bf{r}}_{n}}$. It can be shown that ${\bf \Lambda} = \textrm{diag}(\gamma_1, \gamma_2, \cdots, \gamma_{2K_n})$ with ${\gamma_1} = 2{\gamma_2} =  \cdots  = 2{\gamma_{2{K_n}}} = \sigma _n^2$. Then, in the first step, vector $\bf r_n$  is linearly transformed with unitary matrix $\bf U$ as
	\begin{align} \label{CKLT}
		{\bf{r}}_{nC} = {\bf{U}}^T {\bf{r}}_{n}.
	\end{align}
	The resultant transformed vector ${\bf{r}}_{nC}$ has independent components, each of which is approximately distributed as~\cite{KLT2001}
	\begin{align} \label{distributionofrnC}
		{[{{\bf{r}}_{nC}}]_j} \sim {\mathcal N}({[{{\bf{U}}^T}{{\overline{\bf {r}}}_{n}}]_j},{\gamma_j}),
	\end{align}
	assuming that ${{\bf{r}}_{n}}$ is statistically approximated by ${{\bf{r}'}_{n}}$ as considered in Proposition~\ref{proposition:first:order:approx}.
	
	\par In the second step, a scalar quantization is applied to quantize each component of ${\bf{r}}_{nC}$, subject to a proper bit allocation under the backhaul resource constraint considered. In general, a scalar quantizer with $X_{nj}$-bit resolution is just a mapping $\mathcal{I}: \mathcal{R} \to \mathcal{R}$ whose range has $2^{X_{nj}}$ possible values (forming a codebook). Considering component ${[{{\bf{r}}_{nC}}]_j}$ with distribution $p_L(\cdot) = {\mathcal N}({[{{\bf{U}}^T}{{\overline{\bf {r}}}_{n}}]_j},{\gamma_j})$, we use classic Lloyd quantization~\cite{info_cover} to compress ${[{{\bf{r}}_{nC}}]_j}$  whose codebook can be constructed in an iterative manner as:
	
	\par
	a) Randomly select $2^{X_{nj}}$ real numbers $a_1,\cdots,a_{2^{X_{nj}}}$;
	
	\par
	b) Divide $\mathcal{R}$ into $2^{X_{nj}}$ regions ${\mathcal{R}}_1,\cdots,{\mathcal{R}}_{2^{X_{nj}}}$ with $a_i \in {\mathcal{R}}_i$ and for any $y \in {\mathcal{R}}_i$, $|y-a_i| \le |y-a_k|,1 \le k \le 2^{X_{nj}}, k \ne i$;
	
	\par
	c) Calculate the centroids of these regions as ${m_i} = \frac{{\int_{x \in {{\cal R}_i}} {x{p_L}(x)dx} }}{{\int_{x \in {{\cal R}_i}} {{p_L}(x)dx} }}, i=1,\cdots,2^{X_{nj}}$, and update $a_1,\cdots,a_{2^{X_{nj}}}$ with the centroids $m_1,\cdots,m_{2^{X_{nj}}}$;
	
	\par
	d) Repeat b) and c) until convergence.
	\par It is remarked that each ${[{{\bf{r}}_{nC}}]_j}$ component can be quantized with different $X_{nj}$ bits, however, the total number of bits is constrained by the backhaul resource available as
	\begin{align}
		\sum\nolimits_{j = 1}^{2{K_n}} {{X_{nj}}}  = {C_n}.
	\end{align}
	The bit-allocation optimization for performing the scalar quantization will be further addressed in Sec.~\ref{sec:analysis:opt}.
	
	\subsection{Target Location Estimator at the FC}\label{sec:FCB2}
	\par Assume that the FC can reliably acquire $\hat \tau_n$ and $\hat \alpha_n$, since only two parameters are transmitted and the overhead required is negligible as comparing to transmission of quantized signal samples. As for the signal samples, upon receiving the quantization bits from the $n$-th receiver, the FC looks for the corresponding codeword $[\tilde{\bf{ r}}_{nC}]_j$ from the quantization codebook for the $j$-th component and recovers a compressed signal sample vector ${\bf{\widetilde r}}_{nC}$ for target estimation.
	\par For tractability, it is assumed that each recovered component $[\tilde{\bf{ r}}_{nC}]_j$ is statistically related to the unquantized $[{\bf{ r}}_{nC}]_j$ as
	\begin{align}
		[{{{\bf{\widetilde r}}_{nC}}}]_j = [{{{\bf{r}}_{nC}}}]_j + {q}_{nj},
	\end{align}
	where ${q}_{nj}$ is modeled as an additive Gaussian quantization error with zero mean and variance $\eta_j = \frac{\gamma_j}{2^{2X_{nj}-1}}$ by considering the mutual information
	\begin{align}
		I([{\widetilde{{\bf{r}}}_{nC}}]_j;[{{{\bf{r}}}_{nC}}]_j)= \frac{1}{2}\log_2\left(\frac{\eta_j+\gamma_j}{\eta_j}\right) \le X_{nj}.
	\end{align}
	
	The FC then inversely transforms $\tilde{\bf{ r}}_{nC}$ by multiplying ${\bf{U}}$ and obtain
	\begin{align} \label{decode}
		{\bf{\widetilde r}}_{n} &= {\bf{U}} \tilde{\bf{ r}}_{nC} = {\bf{r}}_{n} + {\bf{U}}{\bf q}_{n},
	\end{align}
	where $\bf U$ is the matrix used in KLT (\ref{CKLT}) and ${\bf{U}}{\bf q}_{n}$ is the effective quantization noise vector $\sim {\mathcal N}({\bf 0},{\bf Q}_n)$, with ${\bf Q}_n$ given by
	\begin{align}\label{Qn}
		{\bf Q}_n={\bf U}\textrm{diag}\left(\left[\eta_1,\cdots,\eta_{2K_n}\right]\right){\bf U}^T.
	\end{align}
	
	\par
	Finally, the target's location vector $\bm{\theta}$ is estimated from $\widetilde{{\bf{r}}}=\left\{{\bf{\widetilde r}}_{n}|n = 1, \cdots ,N\right\}$ and the received estimated values $\hat{{\bm{\tau}}}=\left\{\hat{{ \tau}}_{n}|n = 1, \cdots ,N\right\}$, with $\bm{\alpha}=\left\{\alpha_n|n=1,2,\cdots,N\right\}$ assumed to be perfectly known from their estimated values. Specifically, the ML estimator of $\bm{ \theta}$ is expressed as:
	\begin{align} \label{estiofphi}
		\hat {\bm{\theta}}&= \arg \mathop {\max }\limits_{\bm {\theta}} [\ln p(\widetilde{\bf{r}},\hat{{\bm{\tau}}}|{\bm {\theta}} )] \nonumber \\
		&\mathop  = \limits^{(a)} \arg \mathop {\max }\limits_{\bm {\theta}} [\ln p(\widetilde{\bf{r}}|\hat{{\bm{\tau}}},{\bm {\theta}} ) + \ln p(\hat{{\bm{\tau}}}|{\bm {\theta}} )]
		,
	\end{align}
	where $(a)$ is according to the chain rule of probability theory, and the log-likelihood $\ln p(\widetilde{\bf{r}}|\hat{{\bm{\tau}}},{\bm {\theta}} )$ and $\ln p(\hat{{\bm{\tau}}}|{\bm {\theta}} )$ are given by
	\begin{align} \label{prtheta}
		\ln p(\widetilde{\bf{r}}|\hat{{\bm{\tau}}},{\bm {\theta}} ) &= \nonumber \\
		- \frac{1}{2}&\sum\limits_{n = 1}^N {[{{({{\widetilde {\bf{r}}}_{n}} - {{\bf{s}}_{n}})}}{{\{ {{{\bf{Q}}_{w_n}}} + {{\bf{Q}}_{n}}\} }^{ - 1}}({{\widetilde {\bf{r}}}_{n}} - {{\bf{s}}_{n}})^T]} +  D_1,\nonumber \\
		\ln p({\hat {\bm{\tau}}}|\bm{\theta}) &= - \sum\limits_{n = 1}^N {\frac{1}{2\textrm{CRB}_{\tau _n}}({{\hat \tau }_n} - {\tau _n}(\bm{\theta}))^2} + D_2,
	\end{align}
	where ${{{\bf{Q}}_{w_n}} = \frac{1}{2}\sigma _n^2{\bf I}_{{{{K}}_{n}}}}$ is a $2K_{n}\times2K_{n}$ diagonal matrix,  ${{\bf{s}}_{n}} = \left[\left\{ \sqrt E {\hat \alpha _n}s({k_{n}}{T_s} - {\tau _n}), {k_{n}} \in {{\bf{K}}_{n}} \right\} \right]$ and $D_1$ is a constant independent of $\bm {\theta}$. $D_2$ is a constant uncorrelated with $\bm{\theta}$, since $\hat{\tau} _n\sim {\mathcal N}({\tau _n},\textrm{CRB}_{\tau _n})$ according to~(\ref{Gaussiandelay}).
	\begin{remark}
		In a specific scenario where no backhaul capacity is allocated for quantizing received signal samples, the FC is restricted to estimating the target's location solely based on local estimated values from each receiver. In this case, the ML estimator of $\bm{\theta}$ is given by:
		\begin{align} \label{estiofphi2}
			\hat {\bm {\theta}}  =  \arg \mathop {\max }\limits_{\bm {\theta}} [\ln p(\hat{{\bm{\tau}}}|{\bm {\theta}} )].
		\end{align}
		and we refer to this specific scenario as the baseline design.
	\end{remark}
	
	\section{Optimization for the Proposed Advanced Design} \label{sec:analysis:opt}
	
	\par In this section, we introduce the Expected Cramer-Rao Bound (ECRB) as the optimization metric for the advanced design and formulate an optimization problem to determine the bit allocation for efficient quantization of received signal samples from each receiver, as described in Sec.~\ref{sec:QuantizationB}.
	
	\par Instead of directly considering the ECRB of estimating the target's location, we focus on the ECRB of estimating the time delay $\tau_n$ for better analytical tractability. The prior distribution of $\tau_n$ can be modeled as ${\tau _n} \sim {\mathcal N}({\hat \tau _n}, \textrm{CRB}_{\hat \tau _n})$, where the expression for $\textrm{CRB}_{\hat \tau _n}$ is given by:
	\begin{align}
		\textrm{CRB}_{\hat \tau _n} = \frac{1}{\frac{{2E}}{{\sigma _n^2}}{\left| {{\alpha _n}} \right|^2}\sum\limits_{{k} = 1}^{{K_0}} {{{\left[\frac{{\partial s(t)}}{{\partial t}}|_{t = kT_s  - \hat \tau _n }\right]}^2}}}.
	\end{align}
	Then the ECRB can be expressed as \cite{ZZB1}:
	\begin{align} \label{ECRB}
		\textrm{ECRB} =  \int {\frac{1}{{\sqrt {2\pi \textrm{CRB}_{\hat \tau _n} }}{e^{ - \frac{1}{2 \textrm{CRB}_{\hat \tau _n}}{{({\tau _n} - {{\hat \tau }_n})}^2}}}}\textrm{CRB}'_{\tau _n}} d{\tau _n}
	\end{align}
	where $\textrm{CRB}'_{\tau _n}$  (i.e., the CRB for estimating $\tau_n$ based on the quantized received signal $\widetilde{\bf{r}}$ at the $n$-th receiver) for the advanced design is given by
	\begin{align}
		\textrm{CRB}'_{\tau _n} = \frac{1}{\frac{{\partial {{\bf{s}}_n}}}{{\partial {\tau _n}}}{\{ {{\bf{Q}}_{{w_n}}} + {{\bf{Q}}_n}\} ^{ - 1}}\frac{{\partial {\bf{s}}_n^T}}{{\partial {\tau _n}}}}, \label{CRB}
	\end{align}
	with $\frac{{\partial {{\bf{s}}_n}}}{{\partial {\tau _n}}}=\left[\left\{ {\sqrt E {{\hat \alpha }_n}\frac{{\partial s(t)}}{{\partial t}}{|_{t = {k_n}{T_s} - {\tau _n}}}, {k_n} \in {{\bf{K}}_n}} \right\}\right]$ being the derivative vector.
	
	\par It is noted that the ECRB of \eqref{ECRB} is a function of bit allocation ${X_{nj} ,j=1,\cdots,2K_{n}}$ through ${\bf Q}_n$. Therefore, the following optimization problem can be formulated at the $n$-th receiver:
	\begin{align}
		&\mathop {\min} \limits_{\{X_{nj}\}} \textrm{ECRB} \nonumber \\
		\textrm{s.t.}~&\sum \limits_{j = 1}^{2K_{n}} {X_{nj}} = {C_n}, X_{nj} \in {\mathcal {N}}, j=1,\cdots,2K_{n}. \label{optimization2:constraint}
	\end{align}
	
	\par To solve this problem, we propose a greedy allocation algorithm. Specifically, we initialize all ${X_{nj}}$ to be zero, and then iteratively increase some $X_{nj}$ that can most significantly reduce the ECRB by one bit. We repeat this process until the backhaul resource constraint in~\eqref{optimization2:constraint} is met. The detailed procedures are summarized in Algorithm~\ref{alg:2}.

	\begin{algorithm}\label{alg:1}
		
		\renewcommand{\algorithmicrequire}{\textbf{Input:}}
		
		\renewcommand{\algorithmicensure}{\textbf{Initialization:}}
		
		\caption{An Iterative Greedy Allocation Algorithm}
		
		\begin{algorithmic}[1]\label{alg:2}
			
			\REQUIRE Estimated $\hat \tau_n$ and $\hat \alpha_n$; backhaul capacity $C_n$ of the $n$-th receiver.		
			\ENSURE  ${\textrm{ECRB}_{\textrm{opt}}} \leftarrow \infty$; $X_{nj} \leftarrow 0,j=1,\cdots,2K_{n}$;\\
			Compute ${\bf{Q}}_{{\bf{r}}_{n}}={\bf{U}}\textrm{diag}(\gamma_1,\cdots,\gamma_{2K_{n}}){\bf{U}}^T$ for the given $\hat \tau_n$ and $\hat \alpha_n$.
			\WHILE{$\sum \limits_{j = 1}^{2K_{n}} {X_{nj}} < {C_n}$}
			\FOR{$j=1,\cdots,2K_{1n}$}
			
			\STATE Set $X_{nj} = X_{nj} + 1$;
			\STATE Compute $\eta_j = \frac{\gamma_j}{2^{2X_{nj}}-1},j=1,\cdots,2K_{n}$;
			\STATE Compute ${{\bf{Q}}_n} = {\bf{U}}\textrm{diag}\left(\left[\eta_1^{-1},\cdots,\eta_{2K_{n}}^{-1}\right]\right){\bf{U}}^T$;
			\STATE Compute $\textrm{ECRB}$ according to Equation (\ref{ECRB});
			\IF {$\textrm{ECRB}  \le \textrm{ECRB}_{\textrm{opt}}$}
			\STATE $\textrm{ECRB}_{\textrm{opt}} = \textrm{ECRB} $;
			\STATE $index = j$;
			\ENDIF
			\STATE Set $X_{nj} = X_{nj} - 1$;
			\ENDFOR
			\STATE Set $X_{index} = X_{index} + 1$;
			\ENDWHILE
			\STATE Compute $\eta_j = \frac{\gamma_j}{2^{2X_{nj}}-1},j=1,\cdots,2K_{n}$;
			\STATE Compute ${{\bf{Q}}_n} = {\bf{U}}\left(diag\left(\left[\eta_1^{-1},\cdots,\eta_{2K_{n}}^{-1}\right]\right)\right)^{-1}{\bf{U}}^T$.
		\end{algorithmic}
	\end{algorithm}
	
	\section{Numerical Results}  \label{sec:simulation}
	\begin{figure}[t]
		\centering
		\subfloat[Circular topology]{
			\includegraphics[width=0.22\textwidth]{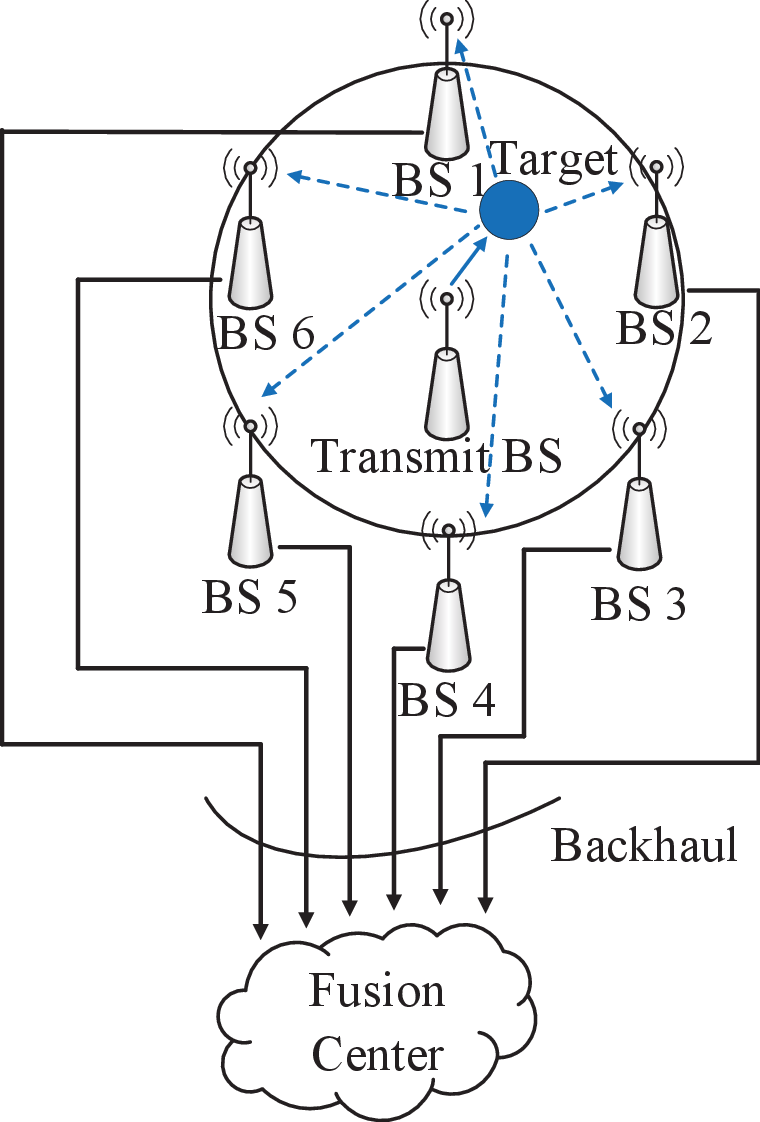}}
		\quad
		\subfloat[Linear topology]{
			\includegraphics[width=0.22\textwidth]{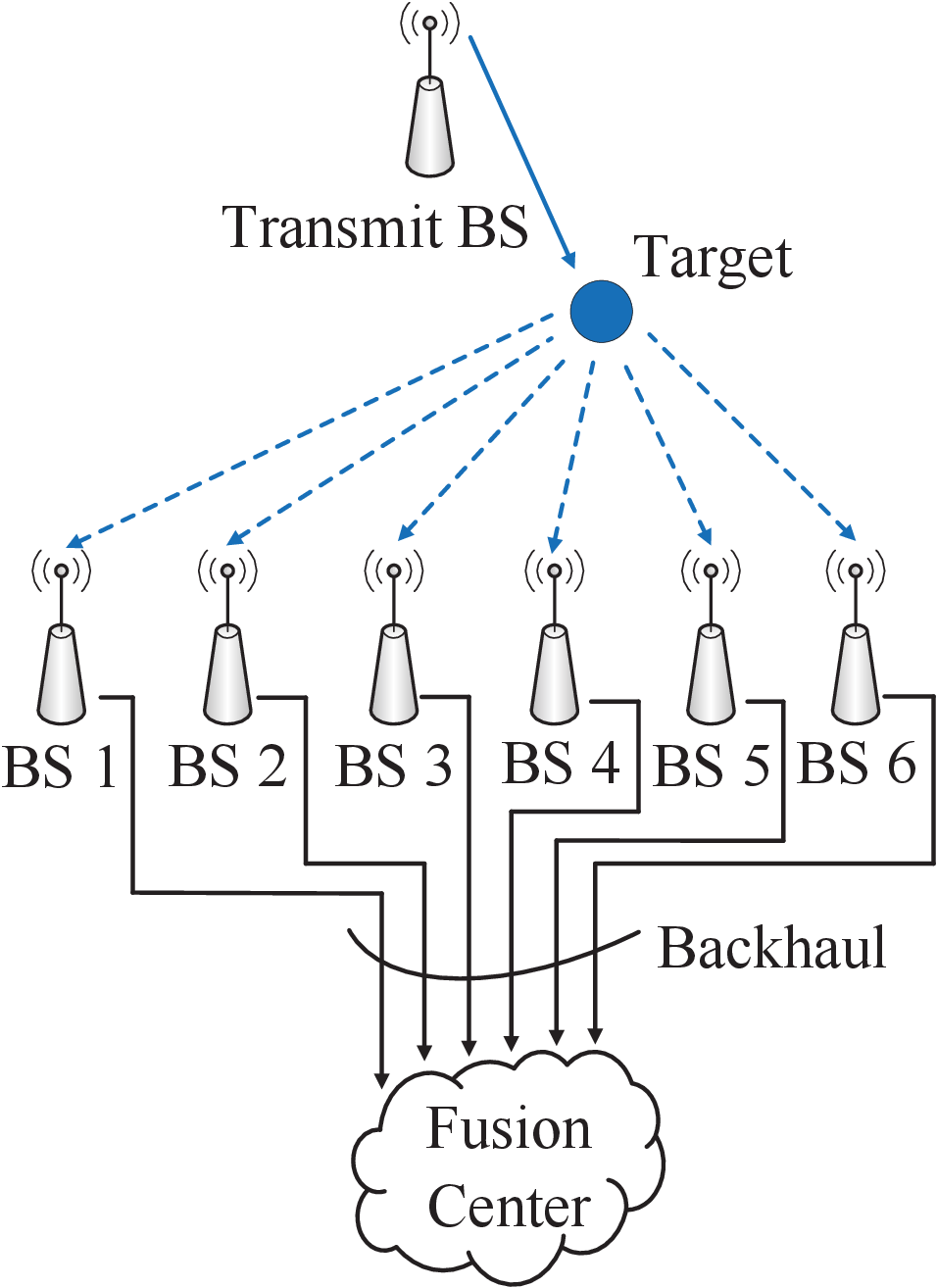}}
		\caption{A case with circular and linear topology of the sensing receivers.}
		\label{Fig:2}
		\vspace*{-1.5\baselineskip}
	\end{figure}
	
	\par Consider a network model comprising one sensing transmitter and $N$ sensing receivers. We simulate two different topologies for the receivers: the circular topology, which represents the optimal distribution of the transmitter and receivers for target estimation \cite{Symmetric}, and the linear topology, suitable for scenarios involving target sensing on roads or railways. As shown in Fig.~\ref{Fig:2}, for a case with $N=6$ sensing receivers, in the circular topology, the receivers are symmetrically distributed on a 500$m$-radius circle around the transmitter. In contrast, for the linear topology, the receivers are linearly distributed on a line with a spacing of 100$m$, and the transmitter is 500$m$ away from the line. Within the circular topology, the target is uniformly distributed within the circle. As for the linear topology, the target is uniformly distributed on a line situated 300$m$ away from the line of receivers.
	
	\par The carrier frequency used is $f_c=3.55$GHz, which corresponds to the centered frequency of the $n78$ frequency band in 5G-NR. We adopt a single-pulse sinusoidal signal as the transmitting signal, with its lowpass equivalent expressed as $s(t) = {(\frac{2}{{{T^2}}})^{0.25}}{e^{( - \pi {t^2}/{T^2})}}$, where $T=2\times10^{-8}s$. The pulse repetition frequency (PRF) is $100$kHz, which means the pulse cycle is $T_p = 10^{-5}s$. The bandwidth is $B=50$MHz, leading to a Nyquist sampling period of $T_{s}=\frac{1}{2B}=10^{-8}s$. Meanwhile, the main lobe of $s(t)$ is in the range $-1.5T\le t \le 1.5T$, resulting in $T_{c}=3T=6\times10^{-8}s$. We take $T_{d}=4T=8\times10^{-8}s$, implying that ${\frac{T_d}{T_{s}}+2=10}$ sampling points are available within one pulse for the advanced design, i.e., $K_{n}=10, n=1,2,\cdots,N$. We assume that the modulus of all reflection coefficients is 1. For simplicity, the same backhaul capacity is adopted for all the receivers, i.e., $C_1 = \cdots = C_N = C$ bits. 
	
	To model the microcell line-of-sight (LoS) path loss, we use the following equation:
	\begin{align}
		L=32.4 + 20\log_{10}(d(km)) + 20\log_{10}(f_c(\textrm{GHz})) \text{ (dB)}, \nonumber
	\end{align}
	where $L(\textrm{dB})$ is the squared pathloss coefficient in dB, and $d(km)$ is the distance between the transmitter and receiver. Additionally, we define the average received Signal-to-Noise Ratio (SNR) of the receivers (RSNR) as:
	\begin{align} \label{RSNR}
		\textrm{RSNR} = 10\log_{10} \sum\nolimits_{n = 1}^N {{{10}^{\frac{{SN{R_n}}}{{10}}}}},
	\end{align}
	where $SNR_{n}(dB)$ represents the received SNR of the $n$-th receiver. The primary performance metric we use to evaluate each design is the MMSE of the target's location. Additionally, we introduce the notion of performance gain (PG) as the ratio of the MMSE obtained by the baseline design to that of the advanced design:
	\begin{align} \label{PG}
		\textrm{PG} = \frac{{\textrm{MMSE}(\textrm{baseline})}}{{\textrm{MMSE}(\textrm{proposed})}}.
	\end{align}

	\begin{figure}[t]
		\centering
		\includegraphics[width=0.35\textwidth]{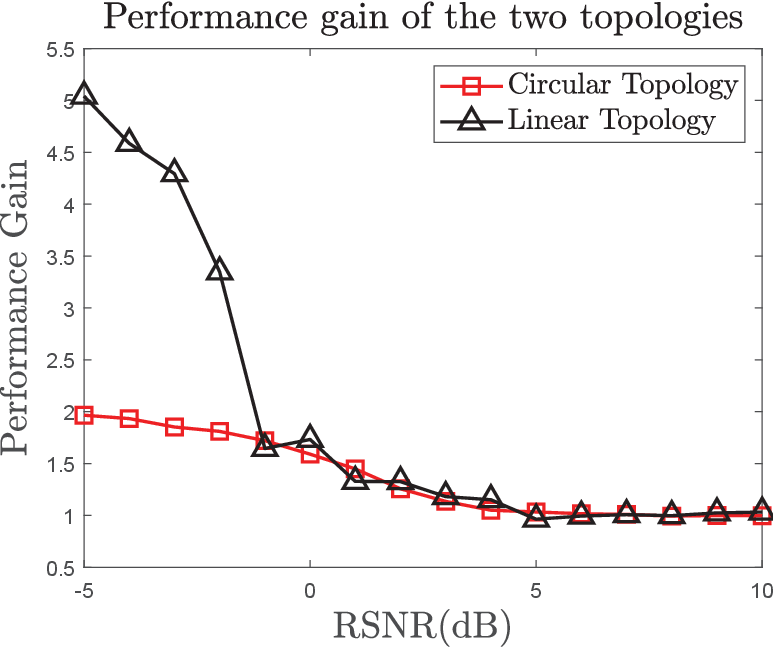}\caption{Performance evaluation for the two topologies considered assuming infinite (sufficiently large) backhaul capacity.}
		\label{Fig:4}
		\vspace*{-1.5\baselineskip}
	\end{figure}
	
	\par Fig.~\ref{Fig:4} presents the performance evaluation for the two considered topologies, assuming infinite (sufficiently large) backhaul capacity. Specifically, the PG defined is plotted as a function of the average RSNR. At medium and high RSNR levels, the PGs of the two topologies are nearly identical. However, for low RSNR values, the PG is more significant in the case of the linear topology compared to that of the circular topology. This difference arises because, at low RSNR, the baseline design's performance in the linear topology degrades significantly more than that of the circular topology.

	\begin{figure}[t]
		\centering
		\subfloat[Circular topology]{
			\includegraphics[width=0.221\textwidth]{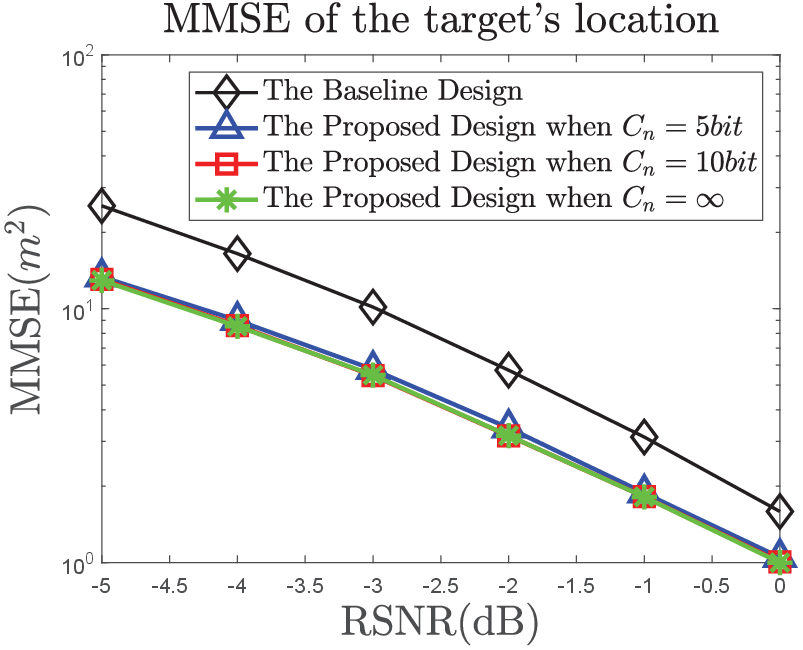}}
		\quad
		\subfloat[Linear topology]{
			\includegraphics[width=0.221\textwidth]{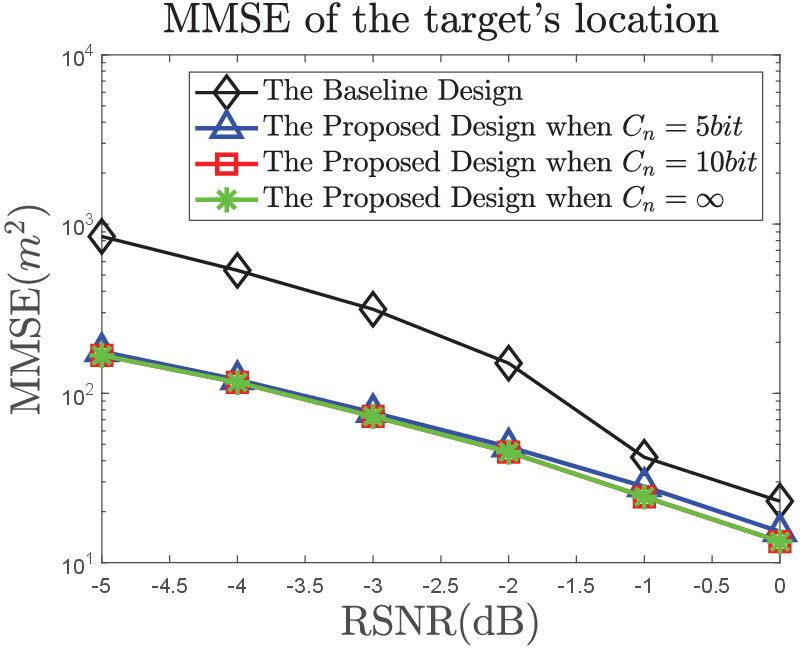}}
		\caption{Comparison between the two designs under different RSNR.}
		\label{Fig:5}
		\vspace*{-1.5\baselineskip}
	\end{figure}
	
	\begin{figure}[t]
		\centering
		\subfloat[Circular topology]{
			\includegraphics[width=0.221\textwidth]{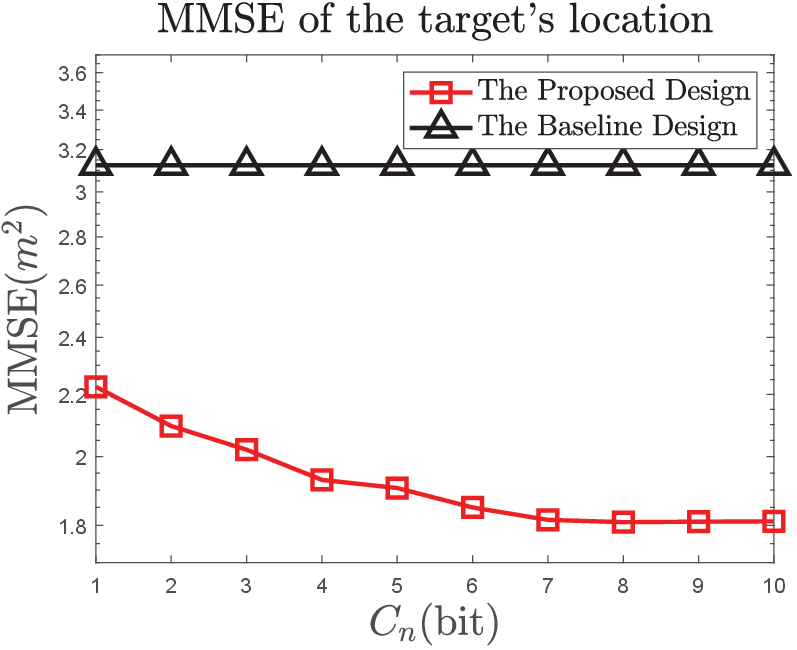}}
		\quad
		\subfloat[Linear topology]{
			\includegraphics[width=0.221\textwidth]{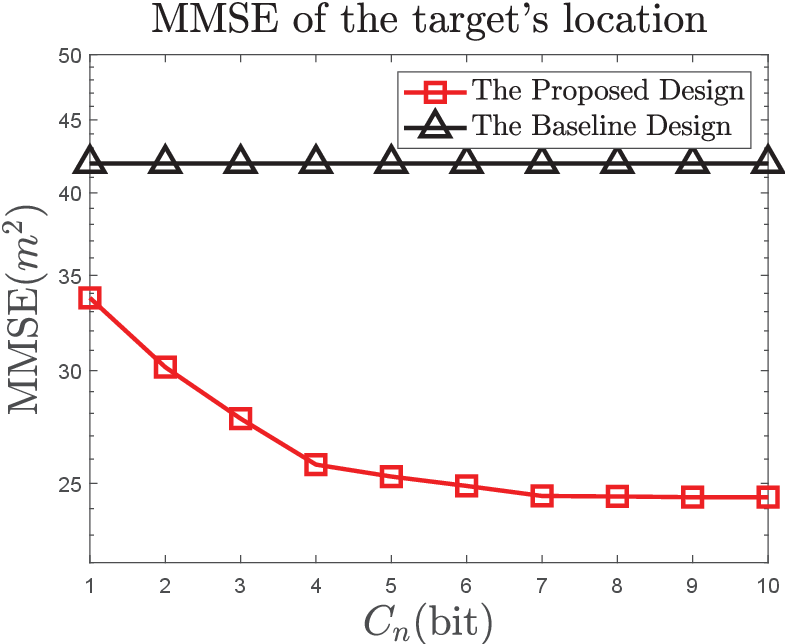}}
		\caption{Comparison between the two designs under different backhaul capacity $C_n$.}
		\label{Fig:6}
		\vspace*{-1.5\baselineskip}
	\end{figure}
	
	\par Fig.~\ref{Fig:5} illustrates the MMSE of the target's location plotted against the RSNR for the two design approaches and the two considered topologies, under different backhaul capacities $C_n$. It is evident that the advanced design consistently outperforms the baseline design across all RSNR values for both topologies. This improvement can be primarily attributed to the additional quantized received signal samples obtained at the FC in the advanced design, which enables more accurate target location estimation. Furthermore, it is worth noting that the performance of the advanced design with $C=10$ bits is remarkably close to the case with infinite bits, resulting in an overhead of $C/T_p = 10/10^{-5} = 1$ Mbs in bit rate for cooperative sensing purpose.  
	
	\begin{figure}[t]
		\centering
		\subfloat[Circular topology]{
			\includegraphics[width=0.221\textwidth]{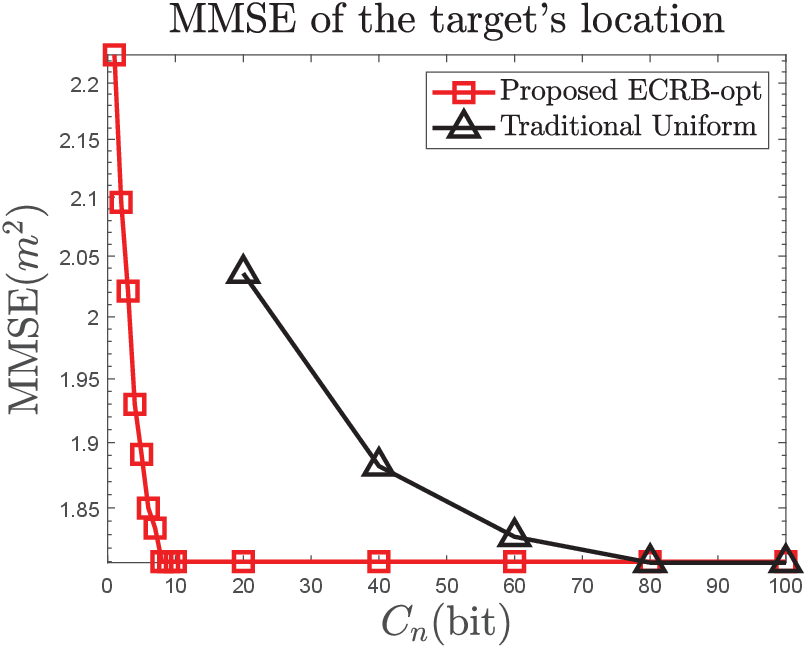}}
		\quad
		\subfloat[Linear topology]{
			\includegraphics[width=0.221\textwidth]{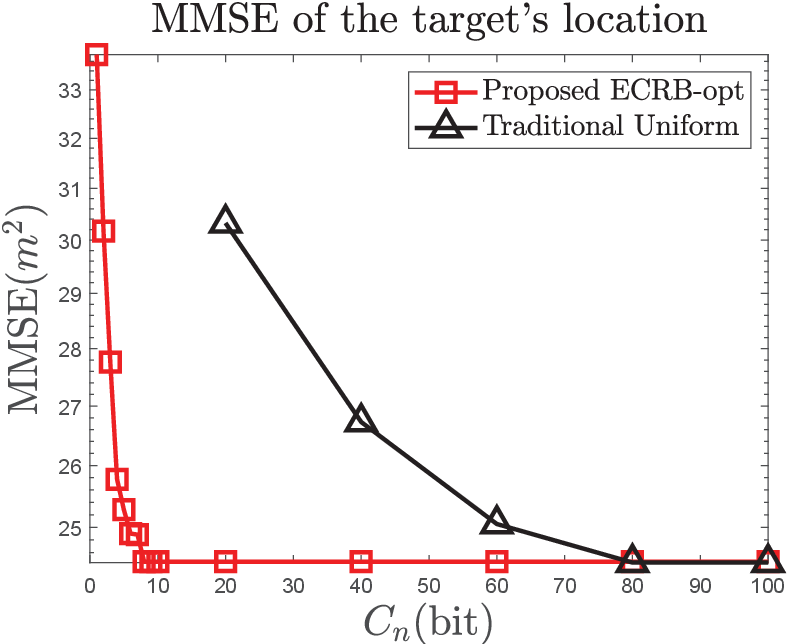}}
		\caption{Comparison between the proposed KLT based quantization and an uniform quantization.}
		\label{Fig:7}
		\vspace*{-1.5\baselineskip}
	\end{figure}
	
	\par With RSNR fixed at $-1$dB, Fig.~\ref{Fig:6} provides a comparison of the MMSE between the two design approaches for the two considered topologies under different backhaul capacities $C_n$. The results show that the advanced design consistently outperforms the baseline design across all backhaul capacities $C_n$ for both topologies, confirming the observations from Fig.~\ref{Fig:5}. In addition, Fig.~\ref{Fig:7} presents the performance achieved by using KLT based quantization and a classic uniform quantization when transmitting the received signal samples to the FC under the two topologies. It is evident that KLT-based quantization outperforms the classic uniform quantization and its advantage becomes apparent when the backhaul capacity is below 80 bits.
	
	\section{Conclusion}\label{sec:conclusion}
	
	\par In this study, we have focused on a cooperative sensing framework in ISAC cellular networks, addressing the challenge of how each sensing receiver processes its local measurements and communicates with the FC via a backhaul link with limited capacity for cooperative target localization. Our proposed advanced design incorporates the delivery of estimated time delay and reflecting coefficient, along with quantized received sensing signals sampled around the local estimated time delay, to the FC using a Karhunen-Lo\`eve Transform coding scheme. To optimize the backhaul resource allocation for quantizing different signal samples, we formulated an optimization problem and introduced an iterative greedy allocation algorithm for solving it. Extensive numerical simulations were conducted to evaluate the performance of the proposed advanced design and demonstrate its superiority over the baseline design across all backhaul capacities. As future work, we are keen to extend the current research to address the challenge of localizing multiple targets simultaneously, which holds significant interest in practice.
	
	\bibliographystyle{IEEETran}

	\small
\end{document}